\documentclass[seceq]{ptptex}

\usepackage{graphicx}
\usepackage{wrapft}

\newcommand{\beq}{\begin{equation}}
\newcommand{\beqa}{\begin{eqnarray}}
\newcommand{\eeq}{\end{equation}}
\newcommand{\eeqa}{\end{eqnarray}}

\newcommand{\bfk}{\boldsymbol{k}}

\newcommand{\bfp}{\boldsymbol{p}}
\newcommand{\bfq}{\boldsymbol{q}}

\notypesetlogo                       

\title{
Third-Order Density Perturbation and
One-Loop Power Spectrum in Dark-Energy-Dominated Universe
}

\author{      
Ryuichi \textsc{Takahashi}
}

\inst{
Department of Physics and Astrophysics, Nagoya University, 
Nagoya 464-8602, Japan 
}

\abst{

We investigate the third-order density perturbation and
 the one-loop correction to the linear power spectrum
 in the dark-energy cosmological model.
Our main interest is to understand the dark-energy effect
 on baryon acoustic oscillations in a
 quasi-nonlinear regime ($k \approx 0.1h$/Mpc).
Analytical solutions and simple fitting formulae are presented for
 the dark-energy model with the general time-varying equation of
 state $w(a)$.
It turns out that the power spectrum coincides with the approximate
 result based on the EdS
 (Einstein de-Sitter) model within $1\%$ for $k<0.4h/$Mpc at $z=0$
 in the WMAP (Wilkinson Microwave Anisotropy Probe) 5yr best-fitting
 cosmological model,
 which suggests that the cosmological dependence is very small.
}

\begin{document}

\maketitle

\section{Introduction}

Revealing the nature of dark energy is fundamentally important 
 not only for astrophysics but also for particle physics.
Constraints on the dark energy from astronomical
 observations is very influential for them.
Baryon acoustic oscillations (BAO) in the galaxy power spectrum
 provide a strong constraint on the dark energy
 using its acoustic scale as a standard ruler.
Large galaxy surveys such as the Sloan Digital Sky Survey and two degree
 field already provide the constraint and future larger surveys are
 currently planned to detect the BAO more accurately\cite{ei05,co05,pe07,ok07}.
Hence, an accurate theoretical model of the BAO is crucial,
and many authors have been investigating the BAO using numerical
 simulation\cite{se05,abfl07,h07,sss07,sss08,t08,sba08}
 and the perturbation theory (including the renormalized
 perturbation theory)\cite{jk06,cs06,cs08,m07,mc07,th08,is07,mp07,n07}. 

Previously, several authors investigated the third-order density
 perturbation and
 derived the one-loop correction to the linear power spectrum in the EdS
 model\cite{j81,j83,ggrw86,ss91,mss92,jb94}.
Similarly for the cosmological constant model,
 Bernardeau (1994) presented the third-order perturbation solution
 (see also Refs. \citen{bjcp92,b94,bchj95,clmm95,m95}).
They found that the dependence of the cosmological model on the second-
 and third-order perturbations is very
 small, if the scale factor in the EdS model is replaced with the
 linear growth factor.\footnote{Martel \& Freudling (1991) and
 Scoccimarro et al. (1998) showed that
 this assumption is valid if $f \equiv d \ln D_1/ d \ln a
 = \Omega_M^{1/2}$. However, since $f \approx \Omega_M^{0.6}$, the
 approximation is not accurate.}
However, since the theoretical model of the BAO should
 archive the subpercent accuracy to provide a strong constraint on the
 dark energy, it is useful to reinvestigate this topic to accurately check
 the above assumption.
In this study, we calculate the third-order density perturbation,
 newly including the dark energy with the time-varying equation of state, 
 and derive the one-loop power spectrum analytically for the first time.
We compare our results with the approximate results based on the EdS model
 in detail, and discuss the effect of the dark energy on the power spectrum
 near the baryon acoustic scale.

Throughout this paper, we use $\delta$ as the density fluctuation,
 $\theta ~(=\nabla \cdot {\boldsymbol{v}})$ as the divergence of the peculiar
 velocity field, and $\tau=a(t) dt$ as the conformal time.
$\Omega_M$, $\Omega_K$ and $\Omega_X$ are the density parameter for
 the matter, the curvature and the dark energy at present.
$w(a)$ is the equation of state of the dark energy.
The Hubble expansion rate is $H^2(a)= H_0^2 \left[ \Omega_M a^{-3} +
 \Omega_K a^{-2} + \Omega_X \exp \left[ 3 \int_a^1 da^\prime
 \left( 1+w(a^\prime) \right)/a^\prime \right] \right]$.

\section{Basics}

The equation of motion determines the growth of the density field
 $\delta(\bfk,\tau)$, and
 velocity field $\theta(\bfk,\tau)$ $(\equiv i \bfk \cdot
 {\boldsymbol{v}}(\bfk, \tau))$
 in the Fourier space is \cite{bcgs02}
\beqa
  && \frac{\partial \delta(\bfk, \tau)}{\partial \tau} + \theta(\bfk,\tau)
 = - \int d^3 \bfq ~\alpha(\bfq,\bfk-\bfq) \theta(\bfq,\tau)
   \delta(\bfk-\bfq,\tau), \label{eom1} \\
 &&  \frac{\partial \theta(\bfk,\tau)}{\partial \tau} + a(\tau) H(\tau)
 \theta(\bfk,\tau)  + \frac{3}{2} a^2(\tau) \Omega_M(\tau) H^2(\tau)
 \delta(\bfk,\tau)  \label{eom2} \nonumber \\
 && ~~~= - \int d^3 \bfq ~\beta(\bfq,\bfk-\bfq) \theta(\bfq,\tau)
   \theta(\bfk-\bfq,\tau),
\eeqa
with 
\beq
   \alpha(\bfp,\bfq)=\frac{\left( \bfp+\bfq \right) \cdot \bfp}{p^2}, ~~
   \beta(\bfp,\bfq)=\frac{\left( \bfp+\bfq \right)^2 \bfp \cdot \bfq}
    {2 p^2 q^2}.
\eeq
Equation (\ref{eom1}) is the continuity equation,
 while equation (\ref{eom2}) is the Euler equation
 with the Poisson equation.
In the linear regime, one can neglect the mode-coupling terms on the
 right-hand sides of Eqs. (\ref{eom1}) and (\ref{eom2}).
Then the linear solutions are
\beq
  \delta_1(\bfk,a)=D_1(a) \delta_1(\bfk), ~\theta_1(\bfk,a)=-a^2 H(a)
 \frac{dD_1(a)}{da} \delta_1(\bfk).
\eeq
The linear growth factor $D_1(a)$ is determined by 
\beq
  \frac{d}{d^2\ln a^2} \frac{D_1}{a} + \left( 4+ \frac{d \ln H}{d \ln a}
 \right) \frac{d}{d\ln a} \frac{D_1}{a} + \left( 3+\frac{d \ln H}{d \ln a}
 - \frac{3}{2} \Omega_M(a) \right) \frac{D_1}{a} = 0,
\eeq
with the initial condition $D_1(a)/a \rightarrow 1$ at $a \rightarrow 0$.
In the special case of the flat model ($\Omega_K=0$) with the constant $w$,
 the solution is given by the hypergeometric function\cite{sw94,p03}.

The density field is formally expanded up to the third order as
 $\delta(\bfk,a)=\delta_1(\bfk,a) +\delta_2(\bfk,a)+\delta_3(\bfk,a)$.
We will show the second- and third-order solutions in the following sections.

\section{Second-order solution}

\begin{figure}[htb]
  \centerline{\includegraphics[width=12.cm]{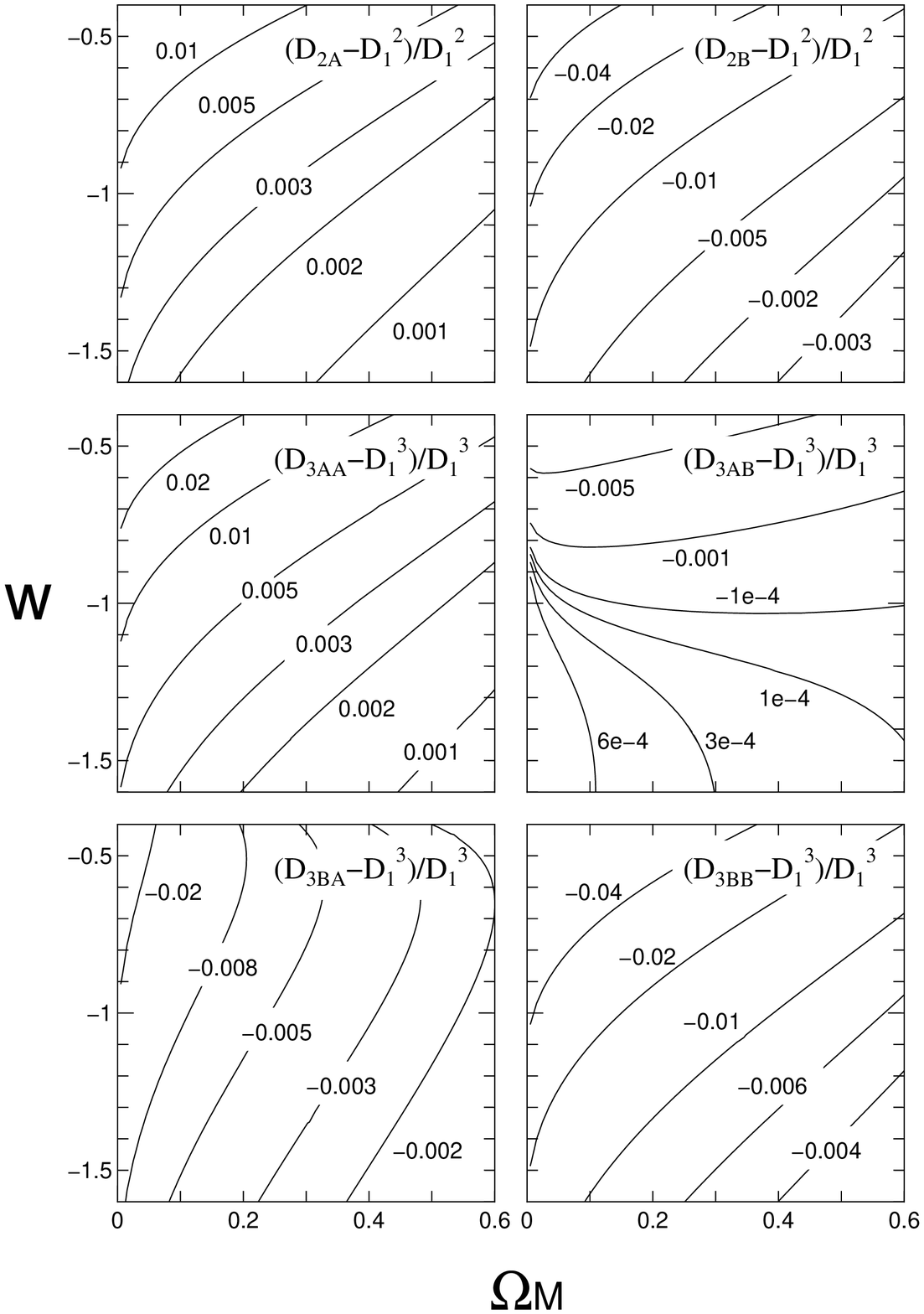}}
  \caption{
Top panels: the contour lines show the relative differences between
  $D_{2 A,B}$ and $D_1^2$ at present ($a=1$) in the $\Omega_M-w$ plane.
The flat cosmological model and the constant equation of state are assumed. 
The top left (right) panel is the result for $D_{2 A}$ ($D_{2 B}$). 
Middle and bottom panels: same as top panels, but for the relative
 differences between $D_{3}$ and $D_1^3$.
}
  \label{fig_f2h3}
\end{figure}

Inserting the linear-order solutions of $\delta_1$ and $\theta_1$ into the
 right-hand sides of equations (\ref{eom1}) and (\ref{eom2}),
 one can obtain the second-order solution as
\beq
  \delta_2(\bfk, a)=D_{2 A}(a) A(\bfk) + D_{2 B}(a) B(\bfk),
\label{d2}
\eeq
with
\beqa
  A(\bfk) = \frac{5}{7} \int d^3 \bfq ~\alpha(\bfq,\bfk-\bfq) \delta_1(\bfq)
  \delta_1(\bfk-\bfq),  \\
  B(\bfk) = \frac{2}{7} \int d^3 \bfq ~\beta(\bfq,\bfk-\bfq) \delta_1(\bfq)
  \delta_1(\bfk-\bfq).
\eeqa
The second-order growth factors $D_{2 A,B}$ are determined by 
 ordinary differential equations with the boundary
 condition $D_{2 A,B} \rightarrow a^2$ at $a \rightarrow 0$
 (see Appendix A).  
One usually approximately use $D_1^2$, instead of $D_{2A,B}$, in
 equation (\ref{d2}).
In order to demonstrate the validity of this approximation,
 we show the relative differences between $D_{2 A,B}$ and $D_1^2$
 in Fig.~\ref{fig_f2h3} for the constant $w$ in the flat model
 ($\Omega_K=0$).
The results are shown by the contour lines in the $\Omega_M - w$ plane
 for $D_{2 A}$ (top left panel) and $D_{2 B}$ (top right panel).
As clearly seen in the figures, the relative errors are
 small, less than $4 \%$ for $0.1<\Omega_M<1$ and $-0.5<w<-1.5$.
The errors become larger for larger $w$.
This tenancy suggests for larger $w$ that the dark energy has been
 affecting the expansion rate since long time ago, and hence the
 large differences between $D_{2A,B}$ and $D_1^2$ arise at present.

\begin{figure}[htb]
  \centerline{\includegraphics[width=12.cm]{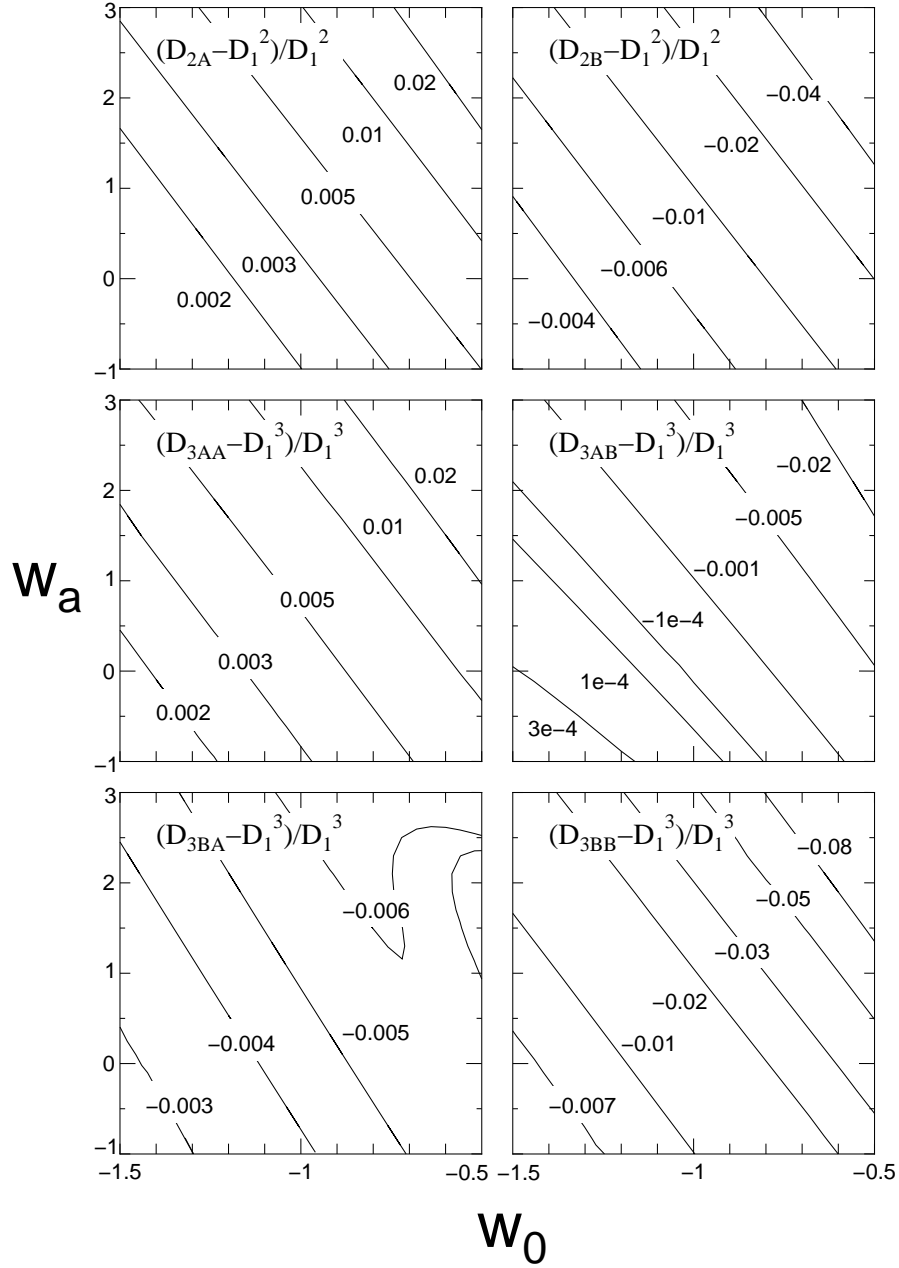}}
  \caption{
Same as Fig.~\ref{fig_f2h3}, but for the time-varying equation of state,
 $w(a)=w_0+w_a a \left( 1-a \right)$.
The results are shown in the $w_0 - w_a$ plane in the flat cosmological
 model with $\Omega_M=0.28$. 
}
  \label{fig_d2d3}
\end{figure}

Figure \ref{fig_d2d3} is the same as Fig.~\ref{fig_f2h3}, but for the time
 varying equation of state\cite{cp01,jbp05}
\beq
  w(a)=w_0+w_a a \left( 1-a \right).
\eeq
The results are shown in the $w_0 - w_a$ plane with
 $\Omega_M=0.28$ $(=1-\Omega_X)$.
As shown in the figure, for large $w_a$, the relative differences
 become large.
This is because the dark energy term in the hubble expansion
 $H^2(a)$, $\Omega_X a^{-3 (1+w_0)}$ $\exp \left[ (3/2)
 w_a (1-a)^2 \right]$, becomes large for large $w_a$ in the past ($a<1$).
The relative errors are less than $10 \%$ for $-1.5<w_0<-0.5$ and
 $w_a<3$.

\section{Third-order solution}

Similarly, the third-order solution consists of six terms, as shown by
\beqa
  \delta_3(\bfk, a)=D_{3 AA}(a) C_{AA}(\bfk)
      + D_{3 AA}^{\prime}(a) C^{\prime}_{AA}(\bfk)
      + D_{3 AB}(a) C_{AB}(\bfk)   \nonumber \\
      + D_{3 AB}^{\prime}(a) C^{\prime}_{AB}(\bfk)
      + D_{3 BA}(a) C_{BA}(\bfk) + D_{3 BB}(a) C_{BB}(\bfk),
\label{d3}
\eeqa
with
\beqa
&&C_{AA}(\bfk) = \frac{7}{18} \int d^3 \bfq ~\alpha(\bfq,\bfk-\bfq)
  \delta_1(\bfq) A(\bfk-\bfq), \\
&&C^{\prime}_{AA}(\bfk) = \frac{7}{30} \int d^3 \bfq ~\alpha(\bfq,\bfk-\bfq)
  \delta_1(\bfk-\bfq) A(\bfq), \\
&&C_{AB}(\bfk) = \frac{7}{18} \int d^3 \bfq ~\alpha(\bfq,\bfk-\bfq)
  \delta_1(\bfq) B(\bfk-\bfq), \\
&&C^{\prime}_{AB}(\bfk) = \frac{7}{9} \int d^3 \bfq ~\alpha(\bfq,\bfk-\bfq)
  \delta_1(\bfk-\bfq) B(\bfq),  \\
&&C_{BA}(\bfk) = \frac{2}{15} \int d^3 \bfq ~\beta(\bfq,\bfk-\bfq)
  \delta_1(\bfq) A(\bfk-\bfq), \\
&&C_{BB}(\bfk) = \frac{4}{9} \int d^3 \bfq ~\beta(\bfq,\bfk-\bfq)
  \delta_1(\bfq) B(\bfk-\bfq).
\eeqa
There are two additional conditions of
\beqa
 \frac{5}{18} D_{3AA} + \frac{2}{9} D_{3AB}^\prime = \frac{1}{2} D_1^3,
 \nonumber \\
 \frac{1}{6} D_{3AA}^\prime + \frac{1}{9} D_{3AB} + \frac{2}{21} D_{3BA}
 + \frac{8}{63} D_{3BB}   = \frac{1}{2} D_1^3, 
\label{d3_cond}
\eeqa
and hence only four terms in Eq.~(\ref{d3}) are independent of each other.
The growth factors $D_{3**}$ are determined by the ordinary
 differential equations with the boundary conditions of
 $D_{3**} \rightarrow a^3$ in $a \rightarrow 0$ (see Appendix A).
The middle and bottom panels in Figs.~\ref{fig_f2h3} and \ref{fig_d2d3}
 are the same as the top panels, but for the relative 
 differences between $D_{3**}$ and $D_1^3$.
The results are shown for $D_{3AA}$ (middle left), $D_{3AB}$
 (middle right), $D_{3BA}$ (bottom left), and $D_{3BB}$ (bottom right). 
The relative differences are less than $7 \%$ for $0.1<\Omega_M<1$ and
 $-1.5<w<-0.5$ and less than $20 \%$ for $-0.5<w_0<0.5$ and $w_a<3$.

Our results of the second- and third-order solutions are consistent with
 the previous results of Bernardeau (1994) for the cosmological constant
 model ($w=-1$).
Although we presented the results for only the density perturbations,
 one can easily obtain the velocity field perturbations
 by inserting Eqs. (\ref{d2}) and (\ref{d3}) to Eqs.
 (\ref{eom1}) and (\ref{eom2}).


\section{One-loop power spectrum}

The one-loop power spectrum is the linear power spectrum with the
 leading correction arising from the second- and third-order density
 perturbations, 
\beqa
 P(k,a) &=& \langle \left| \delta_1(k,a)+\delta_2(k,a)+\delta_3(k,a)
   \right|^2 \rangle \nonumber \\
   &=& D_1^2(a) P_{11}(k) + P_{22}(k,a)+P_{13}(k,a),
\label{one_loop_pk}
\eeqa
where $P_{11}=\langle |\delta_1|^2 \rangle$, $P_{22}=\langle |\delta_2|^2
 \rangle$ and $P_{13}= \langle 2 {\rm{Re}} (\delta_1 \delta_3^*) \rangle$.
The first term is the linear power spectrum, and the second and third
 terms are the one-loop corrections.
The explicit formulae for $P_{22}$ and $P_{13}$ are given in Appendix B.

One usually approximately apply the one-loop power spectrum in the EdS model
 to an arbitrary cosmological model by replacing the scale factor
 by the linear growth factor,
\beq
 P_{\rm EdS}(k,a)=D_1^2(a) P_{11}(k) +  D_1^4(a) \left[ P_{22}(k)+P_{13}(k)
 \right]_{\rm EdS},
\label{one_loop_pk_eds}
\eeq
where the second and third terms are the corrections for the
 EdS model\cite{mss92,jb94} (see also Appendix B). 
We compare the two power spectra in
 Eqs.~(\ref{one_loop_pk}) and (\ref{one_loop_pk_eds}) in order to
 quantitatively demonstrate the validity of the above approximation.
We use CAMB (Code for Anisotropies in the Microwave Background)\cite{lcl00}
 to calculate the linear power spectrum with the cosmological parameters
 $h=0.701$, $\Omega_B=0.0462$, $\Omega_M=0.279$, $n_s=0.96$
 and $\sigma_8=0.82$, consistent with the WMAP 5yr result\cite{k08}.

\begin{figure}[htb]
  \parbox{\halftext}{\includegraphics[width=6.6cm]{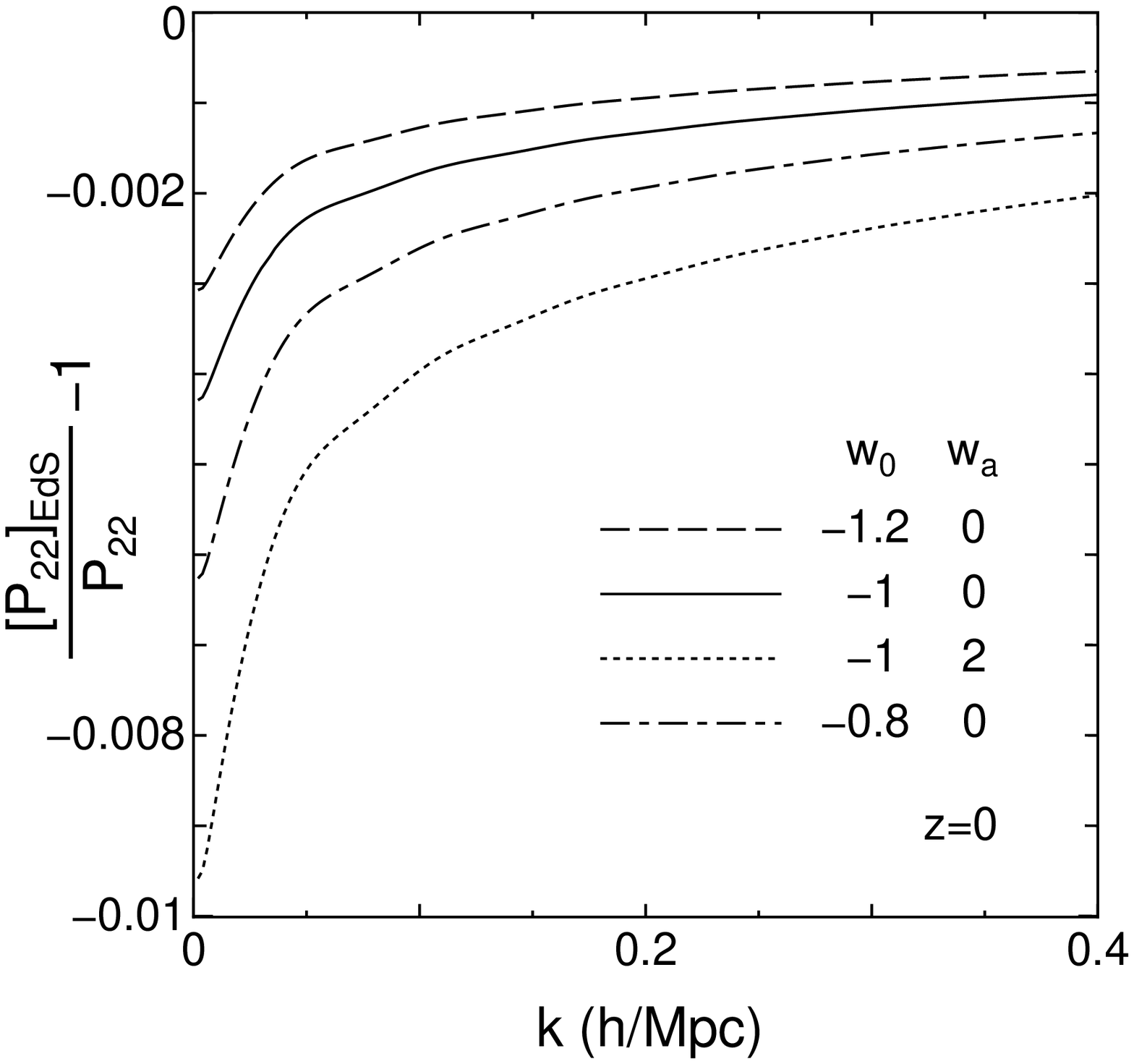}}
  \hspace{0.5cm}
  \parbox{\halftext}{\includegraphics[width=6.6cm]{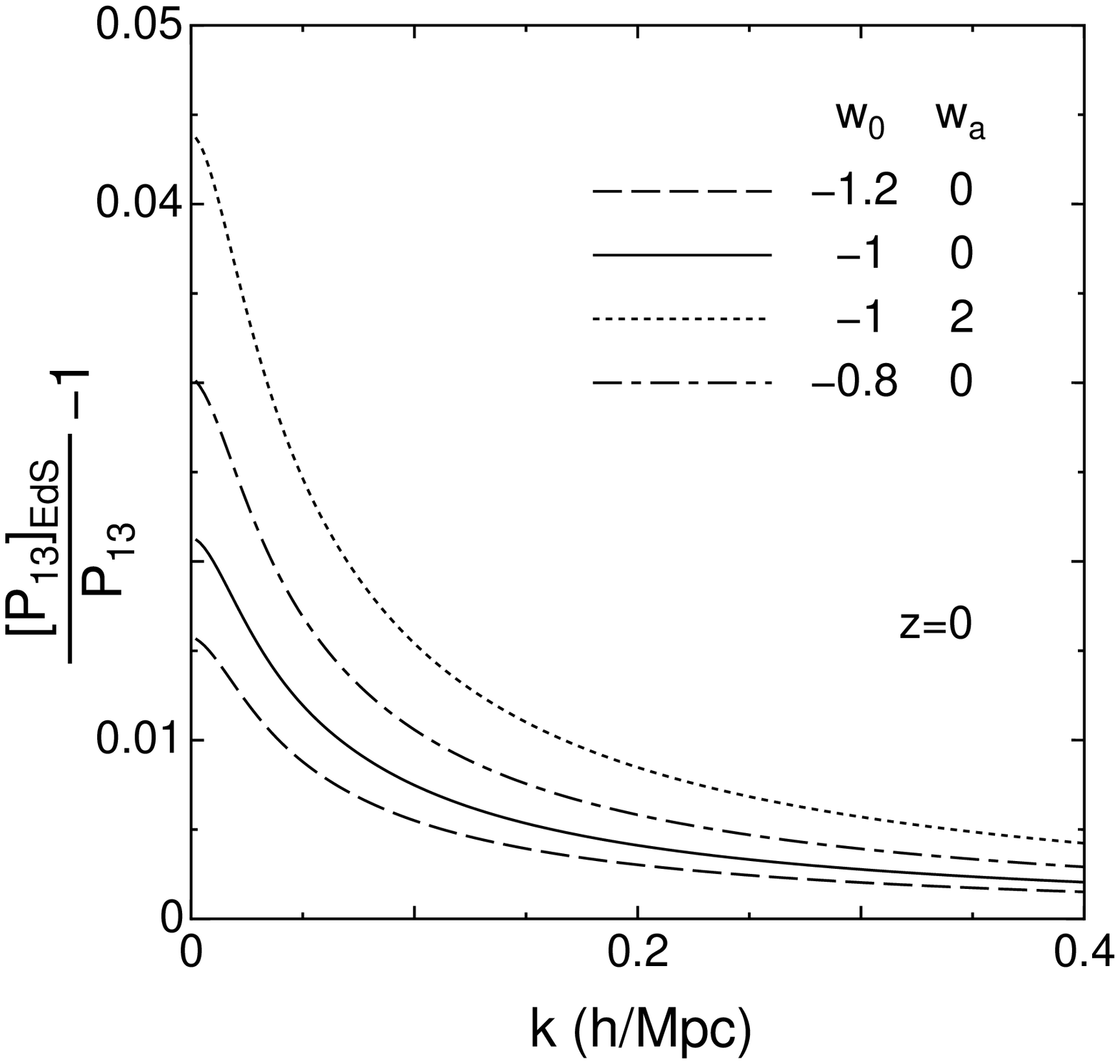}}
  \parbox{\halftext}{\includegraphics[width=6.6cm]{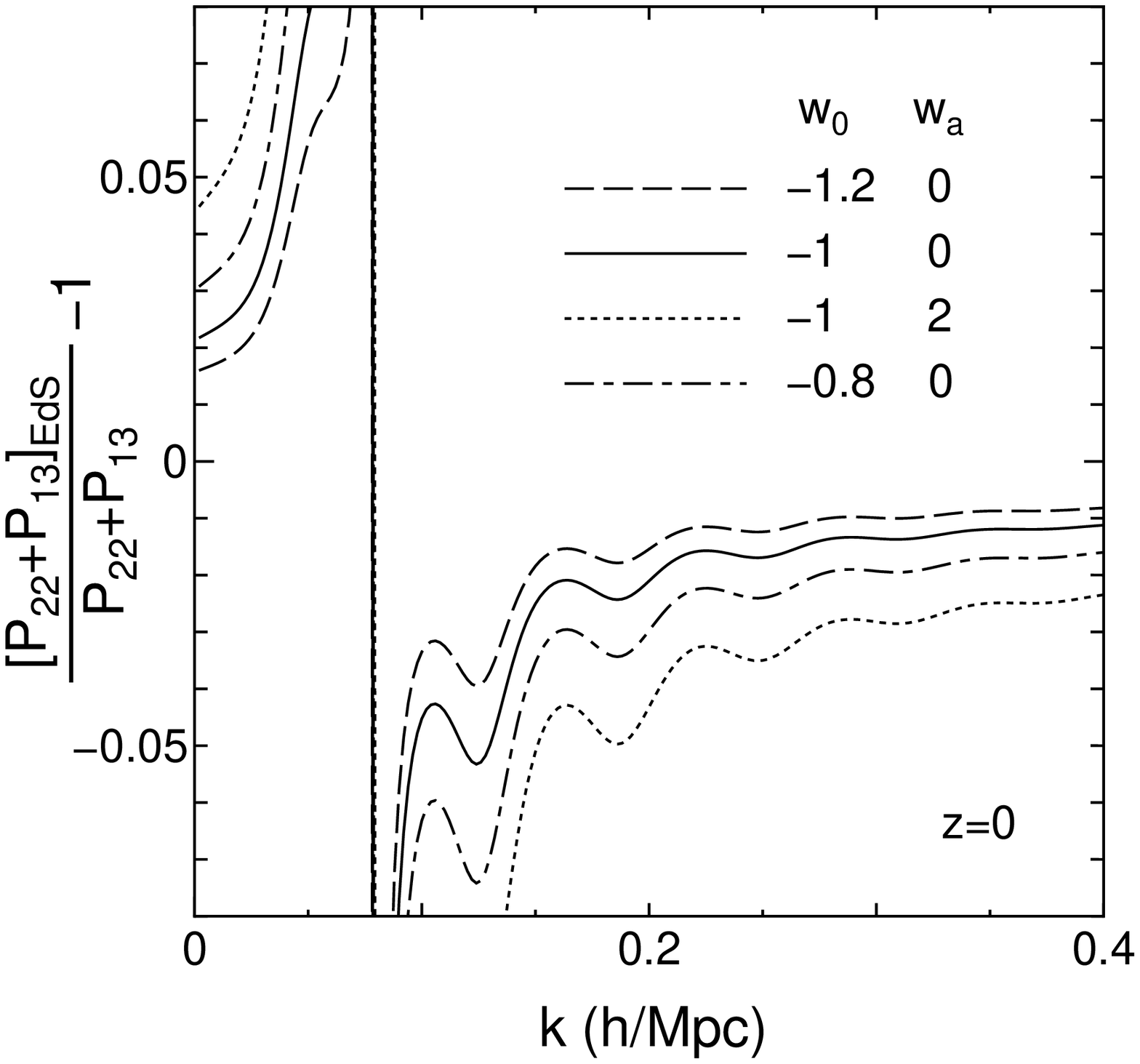}}
  \hspace{0.5cm}
  \parbox{\halftext}{\includegraphics[width=6.6cm]{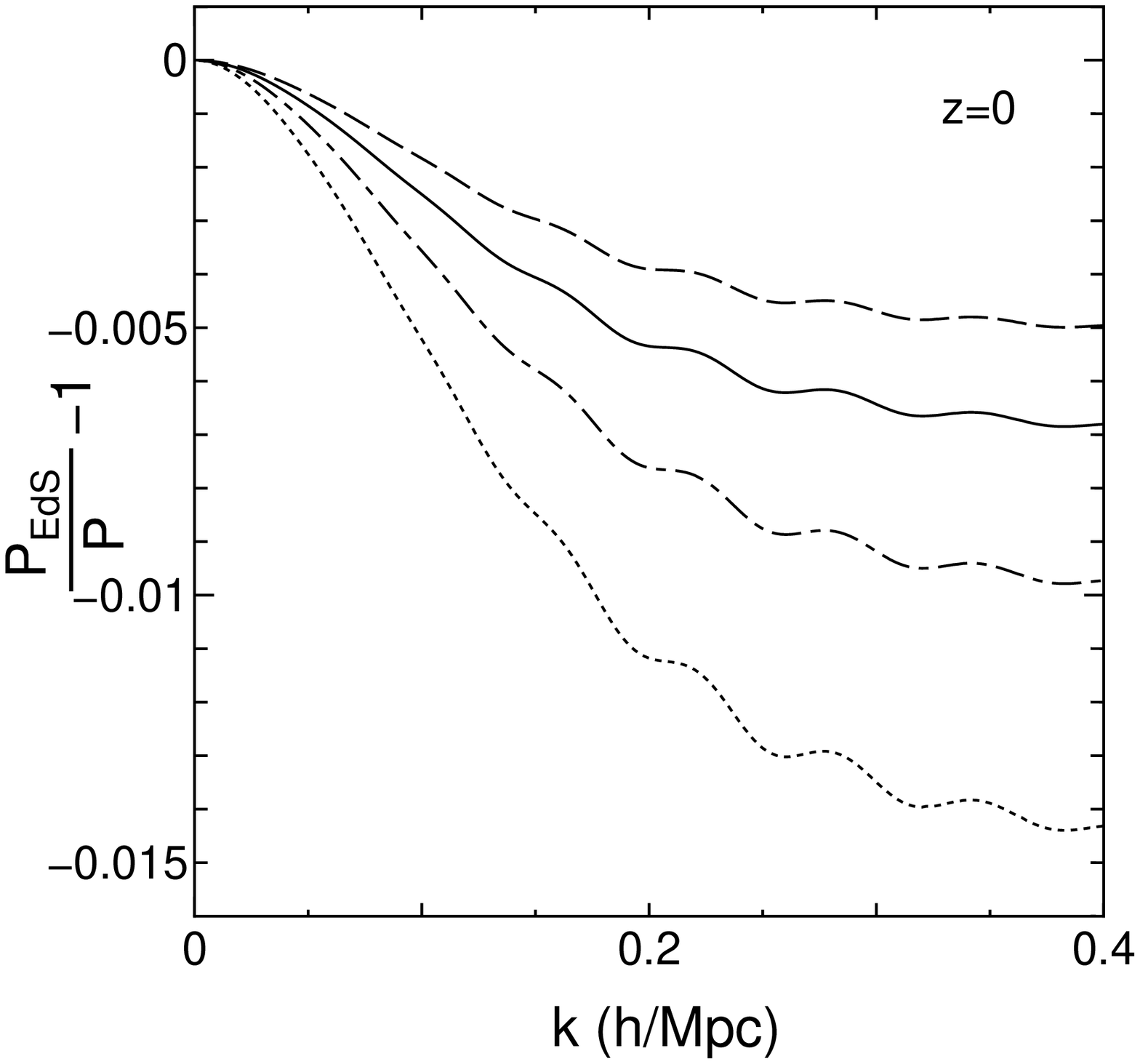}}
  \caption{
Relative differences of $P_{22}(k)$ (top left), $P_{13}(k)$
 (top right), $P_{22}(k)+P_{13}(k)$ (bottom left) and
 $P(k)$ (bottom right) between the correct results and the approximate
 results denoted by $[...]_{\rm EdS}$.
The cosmological model is consistent with the WMAP 5yr result.
}
  \label{fig_dp}
\end{figure}

\begin{figure}[htb]
  \parbox{\halftext}{\includegraphics[width=6.6cm]{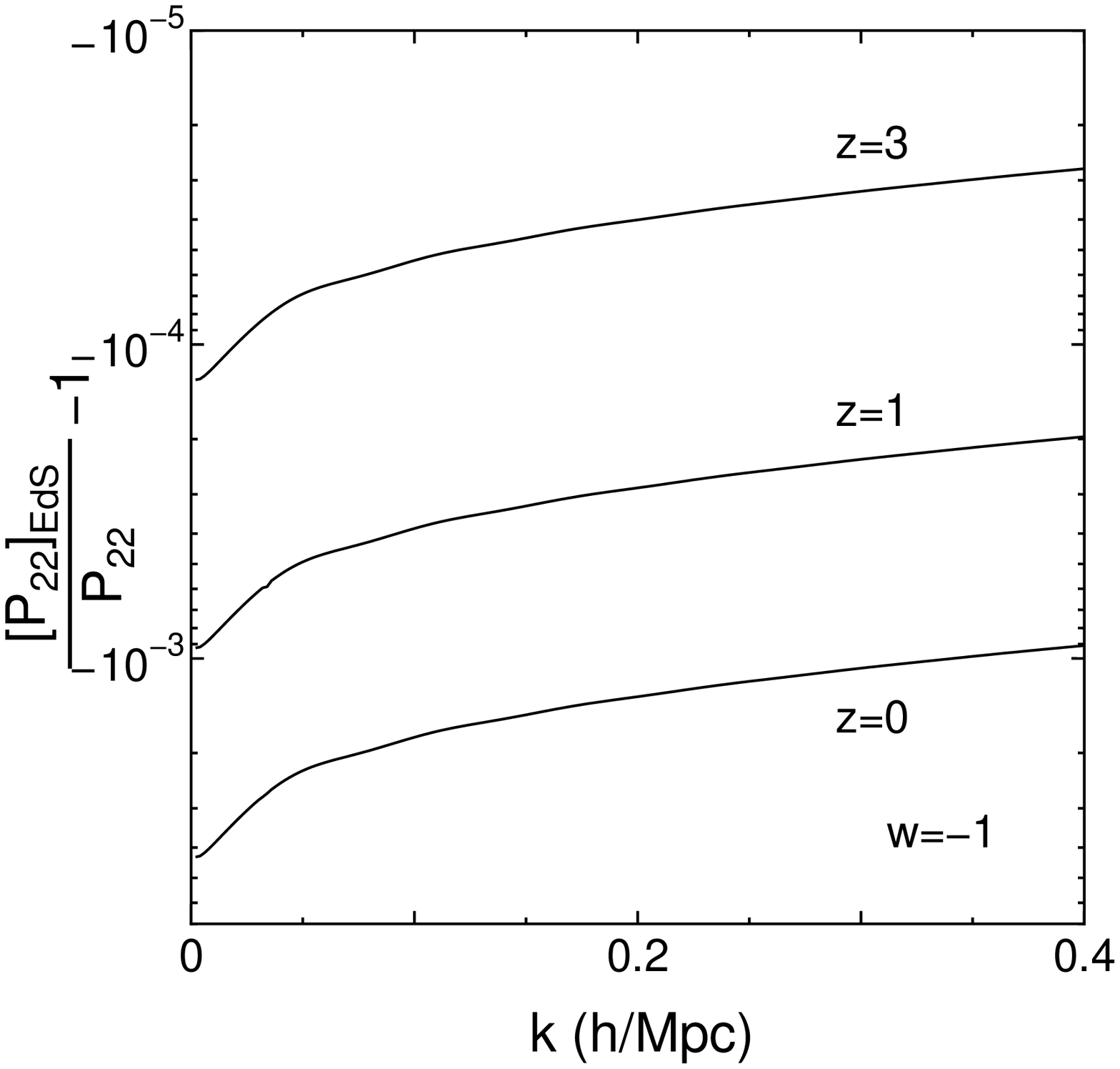}}
  \hspace{0.5cm}
  \parbox{\halftext}{\includegraphics[width=6.6cm]{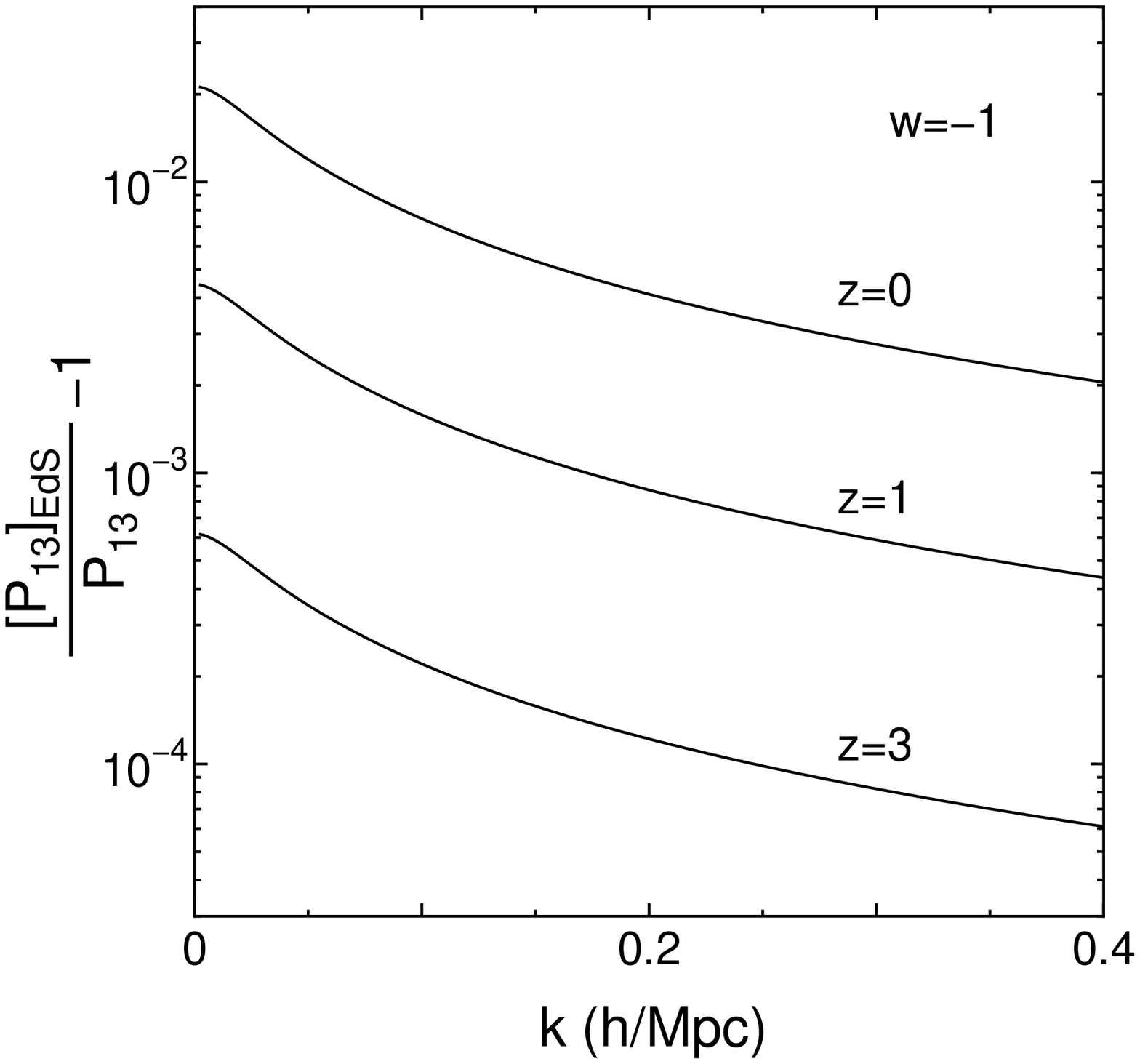}}
  \caption{
Same as top panels in Fig.\ref{fig_dp}, but at various redshifts
 of $z=0,1,3$.
}
  \label{fig_dp_z}
\end{figure}

Figure \ref{fig_dp} shows the relative differences in $P_{22}(k)$,
 $P_{13}(k)$, $P_{22}(k)+P_{13}(k)$ and $P(k)$
 between Eqs.~(\ref{one_loop_pk}) and (\ref{one_loop_pk_eds}) at $z=0$.
The equation of states are $(w_0,w_a)=(-1.2,0)$, $(-1,0)$, $(-1,2)$,
 and $(-0.8,0)$.
From the top panels, the error is $< 5 \%$ for $P_{13}$ while
 $\ll 1 \%$ for $P_{22}$. 
On a small scale, these differences are small.
In the bottom left panel, the error diverges at $k \simeq  0.75h/$Mpc because
 the denominator of $P_{22}+P_{13}$ vanishes there.   
The approximate formula of $P_{\rm EdS}$ predicts a slightly lower value than
 the correct result, 
 because $[P_{22}]_{\rm EdS}(>0)$ is almost the same as $P_{22}$ while
 $[P_{13}]_{\rm EdS}(<0)$ is more negative than $P_{13}$ as shown in the
 top panels.
However, as expected, the difference is very small
 at less than $\sim 1 \%$ for $k<0.4 {\rm h/Mpc}$.
Figure \ref{fig_dp_z} is the same as Fig.~\ref{fig_dp}, but at various
 redshifts of $z=0,1,3$.
Hence, from this figure, the EdS model approximation in
 Eq.~(\ref{one_loop_pk_eds}) is
 sufficiently more accurate for higher redshifts $z>1$.

Finally, we calculate the shift in the position of the first acoustic peak 
 at $k \simeq 0.07h$/Mpc.
Dividing $P(k)$ by the no-wiggle model of Eisenstein \& Hu (1999),
 we find that the position is shifted by only $0.8\% (0.02\%)$ for
 $z=0(z=1)$.

In this chapter, we calculated the one-loop power spectrum,
 however it is not accurate in the strong nonlinear regime
 ($k \gtrsim 0.1h/$Mpc).
In fact, Jeong \& Komatsu (2006) found that the one-loop power
 spectrum coincides with the nonlinear power spectrum from the numerical
 simulation within $1\%$ if $\Delta^2(k)=k^3 P(k)/2\pi^2<0.4$ is satisfied. 
This condition is rewritten as $k<0.12(0.26)h/$Mpc at $z=0(z=1)$.
Hence, in order to extend our analysis to a smaller scale,
 further analysis of the cosmological dependence
 of the higher-order perturbation theory is necessary.

\section{Conclusion}

We investigate the third-order density perturbation
 and the one-loop power spectrum in the dark-energy cosmological model.
We present analytical solutions and a fitting formula
 with the general time-varying equation of state
 for the first time.
It turns out that the cosmological dependence is very weak, 
 for example,
 less than $1 \%$ for $k<0.4h/$Mpc for the power spectrum.
However, our results may be useful in some cases
 when one needs a very highly accurate theoretical model of the BAO
 or in the study of the nonlinear evolution on a smaller scale ($>0.4h$/Mpc).

\section*{Acknowledgements}
We would like to thank Takahiko Matsubara and the anonymous referees
 for helpful comments and suggestions.
This work is supported in part by a 
Grant-in-Aid for Scientific Research on Priority Areas No. 467
``Probing the Dark Energy through an Extremely Wide and Deep Survey with
 Subaru Telescope''.

\appendix
\section{Second- and Third-Order Growth Factors}

The second-order growth factors $D_{2 A,B}$ are determined by
 the ordinary differential equations
\beq
 \frac{d^2}{d \ln a^2} \frac{D_2}{a^2} 
 + \left( 6 + \frac{d \ln H}{d \ln a} \right) \frac{d}{d \ln a}
 \frac{D_2}{a^2}
 + \left[ 8 + 2~\frac{d \ln H}{d \ln a} - \frac{3}{2} \Omega_M(a)
 \right] \frac{D_2}{a^2} \nonumber 
\eeq
\beqa
 &&= \frac{7}{5} \left[ \left( \frac{dD_1}{da} \right)^2 + 
 \frac{3}{2} \Omega_M(a) \left( \frac{D_1}{a} \right)^2 \right]
 ~\mbox{for} ~D_{2A},  \hspace{2cm} \\
 &&= \frac{7}{2}  \left( \frac{dD_1}{da} \right)^2 ~~\mbox{for} ~D_{2B},
 \hspace{2cm}   
\eeqa
with the initial conditions at $a=0$: 
\beq
 \frac{D_{2A,B}}{a^2}=1,~\frac{d}{da} \frac{D_{2A,B}}{a^2}=0.
\eeq
For the flat model with the constant equation of state, 
the solutions are well approximated as
\beqa
  D_{2A} &\simeq& D_1^2 \left[ 1+ \left| \ln \Omega_M \right| \left(
  \frac{5.54 \times 10^{-3}}{|w|} - \frac{3.40 \times 10^{-3}}{\sqrt{|w|}}
 \right) \right],  \\
  D_{2B} &\simeq& D_1^2 \left[ 1+ \left| \ln \Omega_M \right| \left(
  - \frac{1.384 \times 10^{-2}}{|w|} + \frac{8.50 \times 10^{-3}}{\sqrt{|w|}}
 \right) \right],
\eeqa
within a maximum error of $0.03 \%$ for both $0.1 \leq \Omega_M \leq 1$
 and $-1.5 \leq w \leq -0.5$.

\

Similarly, the third-order growth factors are determined by
\beq
 \frac{d^2}{d \ln a^2} \frac{D_3}{a^3} 
 + \left( 8 + \frac{d \ln H}{d \ln a} \right) \frac{d}{d \ln a}
 \frac{D_3}{a^3}
 + \left[ 15 + 3~\frac{d \ln H}{d \ln a} - \frac{3}{2} \Omega_M(a)
 \right] \frac{D_3}{a^3} \nonumber 
\eeq
\beqa
 &&= \frac{18}{7} \left[ 2 \frac{dD_1}{da} +
  \frac{3}{2} \Omega_M(a) \frac{D_1}{a} \right] \frac{D_{2A,B}}{a^2}
  + \frac{18}{7} a \frac{dD_1}{da} \frac{d}{da} \frac{D_{2A,B}}{a^2}
  ~~\mbox{for} ~D_{3AA,3AB}
  \nonumber \\
  \\
 &&= 15 \frac{dD_1}{da} \left[ a \frac{d}{da} \frac{D_{2A}}{a^2} + 2
  \frac{D_{2A}}{a^2} - \frac{7}{5} \frac{D_1}{a} \frac{dD_1}{da}
  \right] ,  ~~\mbox{for} ~D_{3BA} \\
 &&= \frac{9}{2} \frac{dD_1}{da} \left[ a \frac{d}{da} \frac{D_{2B}}{a^2} + 2
  \frac{D_{2B}}{a^2}  \right] ,  ~~\mbox{for} ~D_{3BB} 
\eeqa
with the initial conditions at $a=0$:
\beq
  \frac{D_{3}}{a^3}=1,~\frac{d}{da} \frac{D_{3}}{a^3}=0.
\eeq
For $\Omega_K=0$ with the constant $w$, 
the solutions are well fitted by
\beqa
  D_{3AA} &\simeq& D_1^3 \left[ 1+ \left| \ln \Omega_M \right| \left(
  \frac{8.21 \times 10^{-3}}{|w|} - \frac{5.14 \times 10^{-3}}{\sqrt{|w|}}
 \right) \right],  \\
  D_{3AB} &\simeq& D_1^3 \left[ 1+ \left| \ln \Omega_M \right|^{1.5+0.4 \ln |w|} \left| \Omega_M \right|^{0.7 |w|} \left( - \frac{9.16 \times 10^{-3}}{|w|} + \frac{8.95 \times 10^{-3}}{\sqrt{|w|}}
 \right) \right],  \nonumber \\
  \\
  D_{3BA} &\simeq& D_1^3 \left[ 1+ \left| \ln \Omega_M \right|^{1.06-0.5 \ln |w|} \left( {7.68 \times 10^{-3}}{|w|} - {1.130 \times 10^{-2}}{\sqrt{|w|}}
 \right) \right],  \\
  D_{3BB} &\simeq& D_1^3 \left[ 1+ \left| \ln \Omega_M \right| \left(
 - \frac{2.641 \times 10^{-2}}{|w|} + \frac{1.582 \times 10^{-2}}{\sqrt{|w|}}
 \right) \right],  
\eeqa
within a maximum error of $0.05 \%$ for both $0.1 \leq \Omega_M \leq 1$ and
 $-1.5 \leq w \leq -0.5$.
The other growth factors $D_{3AA}^\prime$ and $D_{3AB}^\prime$ can be
 obtained using Eq.~(\ref{d3_cond}).

\section{Explicit Expressions of $P_{22}$ and $P_{13}$}

Here, we present the explicit expressions of the one-loop correction
 terms $P_{22}$ and $P_{13}$.
From the results in \S $3$, we obtain 
\beq
  P_{22}(k,a)=\langle \left| \delta_2(\bfk,a) \right|^2 \rangle
    = D_{2 A}^2 (a) P_{2 AA}(k) + 2 D_{2 A}(a) D_{2 B}(a) P_{2 AB}(k)
      + D_{2 B}^2 (a) P_{2 BB}(k),
\label{app_p22}
\eeq
with
\beq
 P_{2 AA}(k)= \frac{25}{392 \pi^2} k^4 \int_0^\infty dk_1 \int_{-1}^1 d\mu
  P_{11}(k_1) P_{11} \left( \sqrt{k^2+k_1^2-2 k k_1 \mu} \right) 
  \left( \frac{k\mu+k_1-2k_1\mu^2}{k^2+k_1^2-2kk_1 \mu} \right)^2,
  \nonumber
\eeq
\beq
 P_{2 BB}(k)= \frac{1}{98 \pi^2} k^4 \int_0^\infty dk_1 \int_{-1}^1 d\mu
 P_{11}(k_1) P_{11} \left( \sqrt{k^2+k_1^2-2 k k_1 \mu} \right) 
 \left( \frac{k\mu-k_1}{k^2+k_1^2-2kk_1 \mu} \right)^2,
  \nonumber
\eeq
\beq
 P_{2 AB}(k)= \frac{5}{196 \pi^2} k^4 \int_0^\infty dk_1 \int_{-1}^1 d\mu  
  P_{11}(k_1) P_{11} \left( \sqrt{k^2+k_1^2-2 k k_1 \mu} \right)
 \frac{\left( k\mu-k_1 \right) \left( k\mu+k_1-2k_1\mu^2 \right)}
 {\left( k^2+k_1^2-2kk_1 \mu \right)^2},
 \nonumber
\eeq
where $\mu$ is the cosine between $\bfk$ and $\bfk_1$.

Similarly for $P_{13}$, from the results in \S $4$, we obtain
\beqa
 P_{13}(k,a) &=& \left< 2 {\rm Re} \left( \delta_1(\bfk,a) \delta_3^*(\bfk,a)
      \right) \right> \nonumber \\
      &=& D_{3 AA}(a) P_{3 AA}(k)
      + D_{3 AA}^{\prime}(a) P^{\prime}_{3 AA}(k)
      + D_{3 AB}(a) P_{3 AB}(k)   \nonumber \\
      &+& D_{3 AB}^{\prime}(a) P^{\prime}_{3 AB}(k)
      + D_{3 BA}(a) P_{3 BA}(k) + P_{3 BB}(a) P_{3 BB}(k),
\label{app_p13}
\eeqa
with
\beqa
 &&P_{3 AA}(k)=-\frac{5}{54 \pi^2} k^3 P_{11}(k) \int_0^\infty dr P_{11}(k r)
 \left( 1+r^2 \right),  \nonumber \\
 &&P_{3 AA}^\prime (k)=\frac{1}{24 \pi^2} k^3 P_{11}(k) \int_0^\infty dr
 P_{11}(k r) \left[ 1+4r^2-r^4 + \frac{1}{2r} \left( r^2-1 \right)^3
 \ln \left| \frac{r+1}{r-1} \right| \right],  \nonumber  \\
 &&P_{3 AB} (k)=- \frac{1}{27 \pi^2} k^3 P_{11}(k) \int_0^\infty dr
 P_{11}(k r) \left( 1+r^2 \right), \nonumber  \\
 &&P_{3 AB}^\prime (k)= \frac{2}{27 \pi^2} k^3 P_{11}(k) \int_0^\infty dr
 P_{11}(k r) r^2,  \nonumber  \\
 &&P_{3 BA} (k)=\frac{1}{168 \pi^2} k^3 P_{11}(k) \int_0^\infty dr
 P_{11}(kr) \left[ \frac{2}{r^2} \left( 1-4r^2-r^4 \right) 
 +\frac{(r^2-1)^3}{r^3} \ln \left| \frac{r+1}{r-1} \right| \right],
 \nonumber \\
 &&P_{3 BB} (k)=- \frac{4}{189 \pi^2} k^3 P_{11}(k) \int_0^\infty dr
 P_{11}(kr).  \nonumber  
\eeqa
By setting $D_2=a^2$ and $D_3=a^3$ in Eqs. (\ref{app_p22}) and
 (\ref{app_p13}), the correction terms reduce to
 the result in the EdS model, $a^4[P_{22}(k)+P_{13}(k)]_{\rm EdS}$.

\end{document}